\documentclass[preprint,11pt,showpacs,nofootinbib,floatfix]{revtex4}
\usepackage{amssymb}
\usepackage{amsfonts}
\usepackage{amsmath}
\usepackage{graphicx}
\usepackage{subfigure}
\usepackage[dvipdfm,
            pdfstartview=FitH,
            bookmarksnumbered=true,
            bookmarksopen=true,
            colorlinks,
            pdfborder=001,
            linkcolor=green,
            anchorcolor=green,
            citecolor=red
            ]{hyperref}
\usepackage[toc,page,title,titletoc,header]{appendix}
\begin{document}
\title{Five-dimensional generalized $f(R)$ gravity with curvature-matter coupling}
\author{Ya-Bo Wu}
\thanks{E-mail address:ybwu61@163.com}
\author{Yue-Yue Zhao}
\thanks{zhaoyueyue198737@163.com}
\author{Jun-Wang Lu}
\author{Xue Zhang}
\author{Cheng-Yuan Zhang}
\author{ Jia-Wei Qiao}
\affiliation{
$^{1}$Department of Physics, Liaoning Normal University, Dalian, 116029, China}
\begin{abstract}
The generalized $f(R)$ gravity with curvature-matter coupling in
five-dimensional (5D) spacetime can be established by assuming a
hypersurface-orthogonal spacelike Killing vector field of 5D
spacetime, and it can be reduced to the 4D formulism of FRW
universe. This theory is quite general and can give the
corresponding results to the Einstein gravity, $f(R)$ gravity with
both no-coupling and non-minimal coupling in 5D spacetime as special
cases, that is, we would give the some new results besides previous
ones given by Ref.\cite{60}. Furthermore, in order to get some
insight into the effects of this theory on the 4D spacetime, by
considering a specific type of models with $f_{1}(R)=f_{2}(R)=\alpha
R^{m}$ and $B(L_{m})=L_{m}=-\rho$, we not only discuss the
constraints on the model parameters $m$, $n$, but also illustrate
the evolutionary trajectories of the scale factor $a(t)$, the
deceleration parameter $q(t)$ and the scalar field $\epsilon(t)$,
$\phi(t)$ in the reduced 4D spacetime. The research results show
that this type of $f(R)$ gravity models given by us could explain
the current accelerated expansion of our universe without
introducing dark energy.
\end{abstract}
\maketitle

\section{Introduction}

~~~~As well-known, our current universe is flat and undergoing a phase of the accelerated expansion, which is supported by the recent observational data sets\cite{1,2}. In principle,
this phenomenon can be explained by either dark energy (see Ref.\cite{3} for reviews) which constitutes about three-fourths of the whole matter budget of our universe according to the recent WMAP data\cite{4} and Planck data\cite{5}, in which the reason of this
phenomenon is due to an exotic component with large negative
pressure, or modified theories of gravity\cite{6,7,8,9,10,11}. Alternative to
dark energy, modified theories of gravity is extremely attractive,
such as a new modified gravity theory, namely the so-called $f(T)$ theory, has been proposed recently to drive the current accelerated expansion without invoking dark energy\cite{12,13,14,15,16}. It is a generalized version of the so-called teleparallel gravity originally proposed by Einstein\cite{17}. Moreover, the modified Gauss-Bonnet gravity, i.e., $f(G)$ gravity, where $f(G)$ is a
general function of the Gauss-Bonnet (GB) term, was studied in\cite{18,19,20,21,22}. At
present specific models of $f(G)$ gravity have been proposed to
account for the late-time cosmic acceleration\cite{22,23,24}. Recently the energy conditions
in $f(G)$ gravity have been also discussed\cite{25}, but they are only
adapted to $f(G)$ gravity without coupling between matter and
geometry. $f(G)$ gravity models with
curvature-matter coupling have been proposed, and some relevant
issues, such as the energy conditions, the stability criterion and
the conditions for late-time cosmic accelerated expansion, have been
studied in \cite{26}.

In addition, another interesting alternative modified theory of gravity is $f(R)$ gravity (see, for instance, Ref.\cite{27} for reviews),
here $f(R)$ is an arbitrary function of the Ricci scalar $R$. Cosmic
acceleration can be explained by $f(R)$ gravity\cite{28}, and the
conditions of viable cosmological models have been derived in
\cite{29,30,31,32,33,34}. A general model of $f(R)$ gravity has been proposed in Ref.\cite{35},
which contains a non-minimal coupling between geometry and matter.
This coupling term can be considered as a gravitational source to
explain the current acceleration of the universe. As a result of the
coupling the motion of the massive particles is non-geodesic, and an
extra force, orthogonal to the four-velocity, arises. Different
forms for the Lagrangian density of matter $L_{m}$, and the resulting
extra-force, were considered in \cite{36}, and it was shown that more
natural forms for $L_{m}$ do not imply the vanishing of the
extra-force. The implications of the non-minimal coupling on the
stellar equilibrium were investigated in \cite{37,38}, where
constraints on the coupling were also obtained. An inequality which
expresses a necessary and sufficient condition to avoid the
Dolgov-Kawasaki instability for the model was derived in \cite{39,40,41,42,43,44,45}.
However, a more general model, in which the coupling style is
arbitrary and the Lagrangian density of matter only appears in
coupling term, has been proposed in Ref.\cite{46}, i.e., the
so-called the generalized $f(R)$ gravity with arbitrary coupling
between matter and geometry. In this class of models the
energy-momentum tensor of the matter is generally not conserved, and
the matter-geometry coupling can induce a supplementary acceleration
of the test particles. The first law and the generalized second law of thermodynamics for the generalized $f(R)$ gravity with curvature-matter coupling are studied in the
spatially homogeneous, isotropic FRW universe\cite{47}. Moreover, the energy conditions and  the
Dolgov-Kawasaki criterion for the model have been derived in
Ref.\cite{48,49}, which are quite general and can degenerate to the
well-known energy conditions in GR and $f(R)$ gravity with non-minimal
coupling and non-coupling as special cases.

On the other hand, it is well known that the gravity is the
only dominant long-range interaction, and we have not
fully understood the character of gravity on a cosmological
scale until now. Therefore, the exploration of alternative theories of
gravity were proposed such as the Kaluza-
Klein (KK) theory\cite{50,51,52,53,54}, which was initiated by the motivation
of unifying the gravitation field and the electromagnetic
field in a 5D metric. Since
the fifth dimension is supposed to be a compact $S^{1}$ circle
with an extremely tiny radius, it would actually yield no
observable effect. Recently a new model of modified KK cosmology was studied in \cite{55}, where
the universe in turns inflates, decelerates, and then accelerates
in, early times, radiation dominated
era and matter dominated era, respectively, which shows the results meet the observational
facts roughly. In addition, the Brans-Dicke (BD) theory of gravity
introduced in \cite{56} to match the Mach¡¯s principle, was extended to 5D spacetime, which can naturally predict
the cosmic acceleration without the requirement of a timevarying
BD parameter $\omega$\cite{57} or a fabricated potential in its
4D counterpart\cite{58,59}. Recently, the 5D $f(R)$ theories of
gravity without coupling were studied in \cite{60}, in which the fifth dimension is supposed to be a small unobservable compact ring $S^{1}$ so that a Killing vector field
would arise naturally in the low energy environment. And the 5D theories can be reduced to their 4D formalism by using the Killing reduction. Thus, a natural question we may ask is whether the generalized $f(R)$ gravity with curvature-matter coupling can be extended to 5D spacetime, as well as what effects in the 4D sensational world arise when this 5D generlized $f(R)$ gravity is reduced to the 4D spacetime, which are our motivations and purposes in this paper. The research results show that the extension of the 4D generalized $f(R)$ gravity to 5D spacetime can be realized by assuming a hypersurface-orthogonal spacelike Killing vector field in 5D spacetime, and the evolutionary trajectories of the scale factor $a(t)$, the deceleration parameter $q(t)$ and the scalar field $\epsilon(t)$, $\phi(t)$ of the specific models can be illustrated in the reduced 4D spacetime, which show that this type of models can explain the current accelerated expansion of our universe.

This paper is organized as follows. In Section 2, we will study the
generalized $f(R)$ gravity with curvature-matter coupling in 5D spacetime by using Killing reduction, which can be reduced to a 4D formulism coupled with two scalar fields.
In Section 3, we will discuss the reduced generalized $f(R)$ model in the homogeneous and isotropic universe with the 4D FRW metric. In Section 4, the accelerated universe in the reduced generalized $f(R)$ gravity will be investigated by the numerical analysis on the evolutionary trajectories of the scale factor $a(t)$, the deceleration parameter $q(t)$ and the scalar field $\epsilon(t)$, $\phi(t)$. In the two sections, the corresponding results to the special cases will be discussed. Conclusions of our work will be given in the last section.

\section{Killing reduction of 5D generalized $f(R)$ gravity}
~~~~The action of 5D $f(R)$ gravity \cite{60} is
\begin{equation}\label{1}
S=\int\mathcal{L}^{(5)}\sqrt{-g}d^{5}x=\int[\frac{1}{2\kappa}f(R)+L_{m}]\sqrt{-g}d^{5}x,
\end{equation}
in which $\kappa=8\pi G^{(5)}/c^{4}$, $G^{(5)}$ is the gravitational constant in the 5D spacetime, and $L_{m}$ represents the Lagrangian density of matter.

In the following, we consider the generalized $f(R)$ gravity studied in \cite{46,47,48,49}, in which the coupling style between matter and geometry is arbitrary and the Lagrangian density of matter only appears in coupling term. Now, its action can be extended from 4D to 5D spacetime as follows:
\begin{equation}\label{2}
S=\int\mathcal{L}^{(5)}\sqrt{-g}d^{5}x=\int [\frac{1}{2\kappa}f_{1}(R)+B(L_{m})f_{2}(R)]\sqrt{-g}d^{5}x,
\end{equation}
where $f_{i}(R)(i=1,2)$ and $B(L_{m})$ are arbitrary functions of the Ricci scalar $R$ and the Lagrangian density of matter, respectively. It follows that when $f_{2}(R)=1$ and $B(L_{m})=L_{m}$, Eq.(\ref{2}) can be reduced to Eq.(\ref{1}).

Varying the action (\ref{2}) with respect to the metric $g^{ab}$ of 5D spacetime yields the field equation
\begin{equation}\label{3}
\begin{array}{lcl}
F_{1}(R)R_{ab}-\frac{1}{2}f_{1}(R)g_{ab}+(g_{ab}\Box-\nabla_{a}\nabla_{b})F_{1}(R)
\\=\kappa\{-2B(L_{m})F_{2}(R)R_{ab}-2(g_{ab}\Box-\nabla_{a}\nabla_{b})B(L_{m})F_{2}(R)
\\-f_{2}(R)[K(L_{m})L_{m}-B(L_{m})]g_{ab}+f_{2}(R)K(L_{m})T_{ab}\},
\end{array}
\end{equation}
where $T_{ab}=\frac{-2}{\sqrt{-g}}\frac{\delta S_{m}}{\delta
g^{ab}}$ is the energy-momentum tensor for matter fields,
$\Box=g^{ab}\nabla_{a}\nabla_{b}$, $F_{i}(R)=df_{i}(R)/dR
(i=1,2)$ and $K(L_{m})=dB(L_{m})/dL_{m}$ respectively. Note that $R$, $R_{ab}$ and $T_{ab}$ represent quantities in the 5D spacetime.
Contraction Eq.(\ref{3}) with $g^{ab}$, we obtain the dynamical equation for $F_{1}(R)+2\kappa B(L_{m})F_{2}(R)$:
\begin{equation}\label{4}
\begin{array}{lcl}
\nabla^{a}\nabla_{a}[F_{1}(R)+2\kappa B(L_{m})F_{2}(R)]=\frac{1}{4}\{\frac{5}{2}f_{1}(R)-RF_{1}(R)-2\kappa B(L_{m})F_{2}(R)R\\-5\kappa f_{2}(R)[K(L_{m})L_{m}-B(L_{m})]
+\kappa f_{2}(R)K(L_{m})T^{(5)}\},
\end{array}
\end{equation}
in which $g_{ab}g^{ab}=5$ instead of 4. By using the field equation, the Ricci tensor $R^{(5)}_{ab}$ is given by
\begin{equation}\label{5}
\begin{array}{lcl}
R^{(5)}_{ab}&=&\frac{1}{f'_{1}+2\kappa Bf'_{2}}\{\frac{1}{2}g_{ab}f_{1}-\kappa f_{2}(B'L_{m}-B)g_{ab}+\kappa f_{2}B'T^{(5)}_{ab}-(g_{ab}\Box-\nabla_{a}\nabla_{b})f'_{1}
\\&&-2\kappa (g_{ab}\Box-\nabla_{a}\nabla_{b})(Bf'_{2})\}.
\end{array}
\end{equation}
For convenience, $f_{i}=f_{i}(R)(i=1,2)$, $B=B(L_{m})$ and the prime denotes differentiation with respect to the Ricci scalar $R$ and the Lagrangian density $L_{m}$ respectively.

Below, following the idea of \cite{60}, the structure of the 5D spacetime is considered by assuming
that the 5D spacetime possesses a Killing vector field $\eta^{a}$ which represents the fifth dimension and is everywhere space-like. Hence, the 5D metric can be expressed as
\begin{equation}\label{6}
g_{ab}=h_{ab}+\epsilon^{-1}\eta_{a}\eta_{b},
\end{equation}
where $h_{ab}$ is the metric in the usual 4D universe and $\epsilon=\eta^{a}\eta_{a}$. We choose a coordinate system $\{x^{\mu},x^{5}\}$, $\mu=0,1,2,3$, adapted to the congruence of $\eta^{a}$, i.e., $(\frac{\partial}{\partial x^{5}})^{a}=\eta^{a}$. In the case where $\eta^{a}$ is not hypersurface orthogonal ($\eta_{\mu}\neq0$), $\eta_{\mu}$ will behave as the electromagnetic 4-potential in the reduced 4D theory. Since we are concerned mainly with the cosmological effect of the reduced model, to simplify the discussion, we will consider only the case where $\eta^{a}$ is hypersurface orthogonal. Thus, the line element of $g_{ab}$ reads $ds^{2}=g_{\mu\nu}dx^{\mu}dx^{\nu}+\epsilon dx^{5}dx^{5}$. Through Killing reduction\cite{61,62}, the Ricci tensor $R^{(4)}_{ab}$ of the 4-metric $h_{ab}$ and the scalar field $\epsilon$ are related to the Ricci tensor $R^{(5)}_{ab}$ of $g_{ab}$ by
\begin{equation}\label{7}
R^{(4)}_{ab}=\frac{1}{2}\epsilon^{-1}D_{a}D_{b}\epsilon-\frac{1}{4}\epsilon^{-2}(D_{a}\epsilon)D_{b}\epsilon+h^{m}_{a}h^{n}_{b}R^{(5)}_{mn},
\end{equation}
and
\begin{equation}\label{8}
D^{2}\epsilon=\frac{1}{2}\epsilon^{-1}(D^{a}\epsilon)D_{a}\epsilon-2R^{(5)}_{ab}\eta^{a}\eta^{b},
\end{equation}
where $D_{a}$ is the covariant derivative on 4D spacetime, which satisfies all the conditions of a derivative operator. Furthermore, the stress-energy tensor in (\ref{3}) is regarded as a perfect fluid in 5D spacetime with the form
\begin{equation}\label{9}
T^{(5)}_{ab}=L^{-1}\epsilon^{-\frac{1}{2}}[(\rho+\frac{P}{c^{2}})U_{a}U_{b}+Pg_{ab}]=L^{-1}\epsilon^{-\frac{1}{2}}[T^{(4)}_{ab}
+P\epsilon^{-1}\eta_{a}\eta_{b}],
\end{equation}
in which $\rho$ and $P$ are the 4D energy density and the hydrostatic pressure, respectively, $L$ is the coordinate scale of the fifth dimension. Using the field equation (\ref{3}), Eqs.(\ref{6}) and (\ref{7}), we can obtain the 4D field equation for $h_{ab}$ as
\begin{equation}\label{10}
\begin{array}{lcl}
G^{(4)}_{ab}&=&\frac{1}{2}\epsilon^{-1}(D_{a}D_{b}\epsilon-h_{ab}D^{c}D_{c}\epsilon)-\frac{1}{4}\epsilon^{-2}[(D_{a}\epsilon)D_{b}\epsilon
-h_{ab}(D^{c}\epsilon)D_{c}\epsilon]\\&&+\frac{1}{f'_{1}+2\kappa Bf'_{2}}\{\frac{1}{2}h_{ab}[f_{1}-(f'_{1}+2\kappa Bf'_{2})R-2\kappa f_{2}(B'L_{m}-B)]
+(D_{a}D_{b}-h_{ab}D^{c}D_{c})f'_{1}\\&&+2\kappa(D_{a}D_{b}-h_{ab}D^{c}D_{c})(Bf'_{2})-\frac{1}{2}\epsilon^{-1}h_{ab}(D^{c}\epsilon)D_{c}f'_{1}
-\kappa h_{ab}\epsilon^{-1}(D^{c}\epsilon)D_{c}(Bf'_{2})\\&&+\kappa f_{2}B'L^{-1}\epsilon^{-\frac{1}{2}}T^{(4)}_{ab}\}.
\end{array}
\end{equation}
By using Eq.(\ref{8}), we obtain the dynamical equation of $\epsilon$:
\begin{equation}\label{11}
\begin{array}{lcl}
D^{a}D_{a}\epsilon&=&\frac{1}{2}\epsilon^{-1}(D^{a}\epsilon)D_{a}\epsilon-\frac{1}{f'_{1}+2\kappa Bf'_{2}}\{\frac{\epsilon}{2}[-\frac{1}{2}f_{1}+\kappa f_{2}(B'L_{m}-B)
+(f'_{1}+2\kappa Bf'_{2})R]\\&&+(D^{a}\epsilon)D_{a}f'_{1}+2\kappa(D^{a}\epsilon)D_{a}(Bf'_{2})-\kappa f_{2}B'L^{-1}\epsilon^{\frac{1}{2}}(\frac{1}{2}T^{(4)}-\frac{3}{2}P)\}.
\end{array}
\end{equation}
The dynamical equation of $f'_{1}+2\kappa Bf'_{2}$ is
\begin{equation}\label{12}
\begin{array}{lcl}
D^{a}D_{a}(f'_{1}+2\kappa Bf'_{2})=\frac{1}{4}\{\frac{5}{2}f_{1}-Rf'_{1}-2\kappa Bf'_{2}R-5\kappa f_{2}(B'L_{m}-B)
\\+\kappa f_{2}B'L^{-1}\epsilon^{-\frac{1}{2}}(T^{(4)}+P)\}-\frac{1}{2}\epsilon^{-1}(D^{c}\epsilon)D_{c}f'_{1}
-\kappa\epsilon^{-1}(D^{c}\epsilon)D_{c}(Bf'_{2}).
\end{array}
\end{equation}
The above Eqs.(\ref{10})-(\ref{12}) are just the 4D gravitational field equations.

\section{The reduced generalized $f(R)$ gravity in FRW universe}
~~~~The 4D spatially homogeneous and isotropic FRW universe is considered and described by the metric
\begin{equation}\label{13}
ds^{2}=-dt^{2}+a(t)^{2}(\frac{dr^{2}}{1-kr^{2}}+r^{2}d\Omega^{2}_{2}),
\end{equation}
where $a(t)$ is the scale factor of the universe with $t$ being
cosmic time and $d\Omega^{2}_{2}$ is the metric of
2D sphere with unit radius and the spatial
curvature constant $k$ takes the values $+1,0,-1$ according to a closed, flat and
open universe, respectively. Below, we only consider the spatially flat case, i.e., $k=0$. Thus the two components of the field equation (\ref{10}) are:
\begin{equation}\label{14}
\begin{array}{lcl}
3\left(\frac{\dot{a}}{a}\right)^{2}=-\frac{3}{2}\frac{\dot{a}}{a}\frac{\dot{\epsilon}}{\epsilon}-3\frac{\dot{a}}{a}\frac{\dot{\phi}}{\phi}
-\frac{1}{2}\frac{\dot{\epsilon}}{\epsilon}\frac{\dot{\phi}}{\phi}+\frac{1}{\phi}\left\{\frac{8\pi G}{c^{2}}\cdot f_{2}B'\epsilon^{-\frac{1}{2}}\rho
+\frac{1}{2}\left[2\kappa f_{2}\left(B'L_{m}-B\right)+R\phi-f_{1}\right]\right\},
\end{array}
\end{equation}
and
\begin{equation}\label{15}
\begin{array}{lcl}
-(2\frac{\ddot{a}}{a}+\frac{\dot{a}^{2}}{a^{2}})=\frac{1}{2}\frac{\ddot{\epsilon}}{\epsilon}+\frac{\dot{a}}{a}\frac{\dot{\epsilon}}{\epsilon}
-\frac{1}{4}\frac{\dot{\epsilon}^{2}}{\epsilon^{2}}+\frac{\ddot{\phi}}{\phi}+2\frac{\dot{a}}{a}\frac{\dot{\phi}}{\phi}
+\frac{1}{2}\frac{\dot{\epsilon}}{\epsilon}\frac{\dot{\phi}}{\phi}\\+\frac{1}{2\phi}[f_{1}
-R\phi-2\kappa f_{2}(B'L_{m}-B)]+\frac{8\pi G}{c^{4}}\cdot\frac{f_{2}B'\epsilon^{-\frac{1}{2}}}{\phi}P,
\end{array}
\end{equation}
where $\phi\equiv f'_{1}+2\kappa Bf'_{2}$. Furthermore, the dynamical equations of $\epsilon$, $\phi$ and the scale factor $a(t)$ can be obtained by means of the above equations as well as Eqs.(\ref{11}) and (\ref{12}):
\begin{equation}\label{16}
\begin{array}{lcl}
\frac{\ddot{\epsilon}}{\epsilon}=-3\frac{\dot{a}}{a}\frac{\dot{\epsilon}}{\epsilon}+\frac{1}{2}\frac{\dot{\epsilon}^{2}}{\epsilon^{2}}
-\frac{\dot{\epsilon}}{\epsilon}\frac{\dot{\phi}}{\phi}+\frac{1}{2\phi}[-\frac{1}{2}f_{1}+R\phi+\kappa f_{2}(B'L_{m}-B)]
+\frac{8\pi G}{c^{2}}\cdot\frac{f_{2}B'\epsilon^{-\frac{1}{2}}}{\phi}\cdot\frac{\rho}{2},
\end{array}
\end{equation}
\begin{equation}\label{17}
\begin{array}{lcl}
\frac{\ddot{\phi}}{\phi}=-3\frac{\dot{a}}{a}\frac{\dot{\phi}}{\phi}-\frac{1}{2}\frac{\dot{\epsilon}}{\epsilon}\frac{\dot{\phi}}{\phi}
-\frac{1}{4\phi}[\frac{5}{2}f_{1}-R\phi-5\kappa f_{2}(B'L_{m}-B)]
+\frac{8\pi G}{c^{2}}\cdot\frac{f_{2}B'\epsilon^{-\frac{1}{2}}}{\phi}\cdot(\frac{\rho}{4}-\frac{P}{c^{2}}),
\end{array}
\end{equation}
\begin{equation}\label{18}
\begin{array}{lcl}
\frac{\ddot{a}}{a}=\frac{\dot{a}^{2}}{a^{2}}+\frac{\dot{a}}{a}\frac{\dot{\epsilon}}{\epsilon}+2\frac{\dot{a}}{a}\frac{\dot{\phi}}{\phi}
+\frac{1}{2}\frac{\dot{\epsilon}}{\epsilon}\frac{\dot{\phi}}{\phi}+\frac{1}{4\phi}[\frac{3}{2}f_{1}-R\phi+3\kappa f_{2}(B'L_{m}-B)]-\frac{8\pi G}{c^{2}}\cdot\frac{f_{2}B'\epsilon^{-\frac{1}{2}}}{\phi}\cdot\frac{3\rho}{4}.
\end{array}
\end{equation}
These are the reduced 4D gravitational field equations from the generalized $f(R)$ gravity with curvature-matter coupling in 5D spacetime, which is abbreviated as FGCMC hereafter.

Now some comments on Eqs.(\ref{16})-(\ref{18}) are given as follows:

(1) Let $B(L_{m})=L_{m}$ and $f_{2}(R)$ be rescaled as $1+\lambda f_{2}(R)$, then Eqs.(\ref{16})-(\ref{18}) can be changed into
\begin{equation}\label{19}
\begin{array}{lcl}
\frac{\ddot{\epsilon}}{\epsilon}=-3\frac{\dot{a}}{a}\frac{\dot{\epsilon}}{\epsilon}+\frac{1}{2}\frac{\dot{\epsilon}^{2}}{\epsilon^{2}}
-\frac{\dot{\epsilon}}{\epsilon}\frac{\dot{\phi}}{\phi}+\frac{1}{2\phi}[-\frac{1}{2}f_{1}+R\phi]
+\frac{8\pi G}{c^{2}}\cdot\frac{(1+\lambda f_{2}) \cdot\epsilon^{-\frac{1}{2}}}{\phi}\cdot\frac{\rho}{2},
\end{array}
\end{equation}
\begin{equation}\label{20}
\begin{array}{lcl}
\frac{\ddot{\phi}}{\phi}=-3\frac{\dot{a}}{a}\frac{\dot{\phi}}{\phi}-\frac{1}{2}\frac{\dot{\epsilon}}{\epsilon}\frac{\dot{\phi}}{\phi}
-\frac{1}{4\phi}[\frac{5}{2}f_{1}-R\phi]
+\frac{8\pi G}{c^{2}}\cdot\frac{(1+\lambda f_{2}) \cdot\epsilon^{-\frac{1}{2}}}{\phi}\cdot(\frac{\rho}{4}-\frac{P}{c^{2}}),
\end{array}
\end{equation}
\begin{equation}\label{21}
\begin{array}{lcl}
\frac{\ddot{a}}{a}=\frac{\dot{a}^{2}}{a^{2}}+\frac{\dot{a}}{a}\frac{\dot{\epsilon}}{\epsilon}+2\frac{\dot{a}}{a}\frac{\dot{\phi}}{\phi}
+\frac{1}{2}\frac{\dot{\epsilon}}{\epsilon}\frac{\dot{\phi}}{\phi}+\frac{1}{4\phi}[\frac{3}{2}f_{1}-R\phi]-\frac{8\pi G}{c^{2}}\cdot\frac{(1+\lambda f_{2}) \cdot\epsilon^{-\frac{1}{2}}}{\phi}\cdot\frac{3\rho}{4}.
\end{array}
\end{equation}
Here $\phi=f'_{1}+2\kappa\lambda L_{m}f'_{2}$. Evidently, Eqs.(\ref{19})-(\ref{21}) are just the reduced 4D gravitational field equations from the 5D $f(R)$ gravity with non-minimal coupling, FGNMC for short.

(2)By setting $B(L_{m})=L_{m}$, $f_{2}(R)=1$, which results in $\phi=f'_{1}$, in the case the corresponding results to Eqs.(\ref{16})-(\ref{18}) are as follows:
\begin{equation}\label{22}
\begin{array}{lcl}
\frac{\ddot{\epsilon}}{\epsilon}=-3\frac{\dot{a}}{a}\frac{\dot{\epsilon}}{\epsilon}+\frac{1}{2}\frac{\dot{\epsilon}^{2}}{\epsilon^{2}}
-\frac{\dot{\epsilon}}{\epsilon}\frac{\dot{\phi}}{\phi}+\frac{1}{2\phi}[-\frac{1}{2}f_{1}+R\phi]
+\frac{8\pi G}{c^{2}}\cdot\frac{\epsilon^{-\frac{1}{2}}}{\phi}\cdot\frac{\rho}{2},
\end{array}
\end{equation}
\begin{equation}\label{23}
\begin{array}{lcl}
\frac{\ddot{\phi}}{\phi}=-3\frac{\dot{a}}{a}\frac{\dot{\phi}}{\phi}-\frac{1}{2}\frac{\dot{\epsilon}}{\epsilon}\frac{\dot{\phi}}{\phi}
-\frac{1}{4\phi}[\frac{5}{2}f_{1}-R\phi]
+\frac{8\pi G}{c^{2}}\cdot\frac{\epsilon^{-\frac{1}{2}}}{\phi}\cdot(\frac{\rho}{4}-\frac{P}{c^{2}}),
\end{array}
\end{equation}
\begin{equation}\label{24}
\begin{array}{lcl}
\frac{\ddot{a}}{a}=\frac{\dot{a}^{2}}{a^{2}}+\frac{\dot{a}}{a}\frac{\dot{\epsilon}}{\epsilon}+2\frac{\dot{a}}{a}\frac{\dot{\phi}}{\phi}
+\frac{1}{2}\frac{\dot{\epsilon}}{\epsilon}\frac{\dot{\phi}}{\phi}+\frac{1}{4\phi}[\frac{3}{2}f_{1}-R\phi]-\frac{8\pi G}{c^{2}}\cdot\frac{\epsilon^{-\frac{1}{2}}}{\phi}\cdot\frac{3\rho}{4}.
\end{array}
\end{equation}
The above equations are just reduced ones from the 5D pure $f(R)$ gravity with no-coupling, FGNC for short, which are consistent with the ones in Ref.\cite{60}.

(3)If $B(L_{m})=L_{m}$, $f_{2}(R)=1$ and $f_{1}(R)=R$, we have $\phi=1$. Thus, Eqs.(\ref{16})-(\ref{18}) become
\begin{equation}\label{25}
\begin{array}{lcl}
\frac{\ddot{\epsilon}}{\epsilon}=-3\frac{\dot{a}}{a}\frac{\dot{\epsilon}}{\epsilon}+\frac{1}{2}\frac{\dot{\epsilon}^{2}}{\epsilon^{2}}
+\frac{R}{4}
+\frac{8\pi G}{c^{2}}\cdot\frac{\epsilon^{-\frac{1}{2}}\rho}{2},
\end{array}
\end{equation}
\begin{equation}\label{26}
\begin{array}{lcl}
\frac{\ddot{a}}{a}=\frac{\dot{a}^{2}}{a^{2}}+\frac{\dot{a}}{a}\frac{\dot{\epsilon}}{\epsilon}
+\frac{R}{8}-\frac{8\pi G}{c^{2}}\cdot\frac{3\epsilon^{-\frac{1}{2}}\rho}{4},
\end{array}
\end{equation}
which are just the reduced 4D gravitational field equations from the 5D Einstein gravity (EG).

It follows that Eqs.(\ref{16})-(\ref{18}) are quite general and can
give the corresponding results to the Einstein gravity, $f(R)$
gravity with no-coupling and non-minimal coupling in 5D spacetime as
special cases.

\section{The accelerated universe for the reduced generalized $f(R)$ gravity}
~~~~Now, we consider a specific type of $f(R)$ models,
$f_{1}(R)=f_{2}(R)=\alpha R^{m}$ ($m\neq1$), and choose
$B(L_{m})=L_{m}=-\rho$ in order to further study the evolutional
characters of some cosmological quantities such as the scale factor
$a(t)$, the deceleration parameter $q(t)$ etc. According to $
\phi\equiv f'_{1}+2\kappa Bf'_{2}=\alpha m(1-2\kappa\rho)R^{m-1}$,
we can obtain $R=[\frac{\phi}{\alpha
m(1-2\kappa\rho)}]^\frac{1}{m-1}$,
$f_{1}(R)=f_{2}(R)=\frac{\phi}{m(1-2\kappa\rho)}[\frac{\phi}{\alpha
m(1-2\kappa\rho)}]^\frac{1}{m-1}$, where $\alpha$ is a dimensional
constant. Furthermore, we consider the present epoch of the universe
is matter-dominated and choose $p=0$ and
$\rho=\rho_{0}(\frac{a_{0}}{a})^{3}$, thus the three cosmological
evolution equations from the combination of
Eqs.(\ref{16})-(\ref{18}) are:
\begin{equation}\label{27}
\begin{array}{lcl}
\frac{\ddot{\epsilon}}{\epsilon}=-3\frac{\dot{a}}{a}\frac{\dot{\epsilon}}{\epsilon}+\frac{1}{2}\frac{\dot{\epsilon}^{2}}{\epsilon^{2}}
-\frac{\dot{\epsilon}}{\epsilon}\frac{\dot{\phi}}{\phi}-\left[\frac{\phi}{\alpha m\left(1-2\kappa\rho_{0}\left(\frac{a_{0}}{a}\right)^{3}\right)}\right]^\frac{1}{m-1}\left[\frac{1-2m\left(1-2\kappa\rho_{0}\left
(\frac{a_{0}}{a}\right)^{3}\right)
-2\epsilon^{-\frac{1}{2}}\cdot c^{2}\kappa L^{-1}\cdot\rho_{0}\left(\frac{a_{0}}{a}\right)^{3}}{4m\left(1-2\kappa\rho_{0}\left(\frac{a_{0}}{a}\right)^{3}\right)}\right],
\end{array}
\end{equation}
\begin{equation}\label{28}
\begin{array}{lcl}
\frac{\ddot{\phi}}{\phi}=-3\frac{\dot{a}}{a}\frac{\dot{\phi}}{\phi}-\frac{1}{2}\frac{\dot{\epsilon}}{\epsilon}\frac{\dot{\phi}}{\phi}
-\left[\frac{\phi}{\alpha m\left(1-2\kappa\rho_{0}\left(\frac{a_{0}}{a}\right)^{3}\right)}\right]^\frac{1}{m-1}\left[\frac{5-2m\left(1-2\kappa\rho_{0}\left
(\frac{a_{0}}{a}\right)^{3}\right)
-2\epsilon^{-\frac{1}{2}}\cdot c^{2}\kappa L^{-1}\cdot\rho_{0}\left(\frac{a_{0}}{a}\right)^{3}}{8m\left(1-2\kappa\rho_{0}\left(\frac{a_{0}}{a}\right)^{3}\right)}\right],
\end{array}
\end{equation}
\begin{equation}\label{29}
\begin{array}{lcl}
\frac{\ddot{a}}{a}=\left(\frac{\dot{a}}{a}\right)^{2}+\frac{\dot{a}}{a}\frac{\dot{\epsilon}}{\epsilon}+2\frac{\dot{a}}{a}\frac{\dot{\phi}}{\phi}
+\frac{1}{2}\frac{\dot{\epsilon}}{\epsilon}\frac{\dot{\phi}}{\phi}+\left[\frac{\phi}{\alpha m\left(1-2\kappa\rho_{0}\left(\frac{a_{0}}{a}\right)^{3}\right)}\right]^\frac{1}{m-1}\left[\frac{3-2m\left(1-2\kappa\rho_{0}
\left(\frac{a_{0}}{a}\right)^{3}\right)
-6\epsilon^{-\frac{1}{2}}\cdot c^{2}\kappa L^{-1}\cdot\rho_{0}\left(\frac{a_{0}}{a}\right)^{3}}{8m\left(1-2\kappa\rho_{0}\left(\frac{a_{0}}{a}\right)^{3}\right)}\right].
\end{array}
\end{equation}

It is worth stressing that for the above derivations we have taken
$G=6.67\times10^{-11}kg^{-1}m^{3}s^{-2}$,
$\rho_{0}=(3.8\pm0.2)\times10^{-28}kgm^{-3}$,
$H_{0}=(2.3\pm0.1)\times10^{-18}s^{-1}$\cite{63} and $G^{(5)}=G\cdot
L=6.67\times10^{-44}kg^{-1}m^{4}s^{-2}$ (note $L=10^{-34}m$). For
the numerical simulation, $a_{0}$, $\epsilon_{0}$ themselves have no
direct physical meaning, therefore they can be simply fixed as 1
with no dimension. Thus, we can directly determine the value of
$\dot{a}_{0}$ by the present value of Hubble parameter
$H_{0}=(\frac{\dot{a}}{a})_{t_{0}}$. Furthermore, we adopt the idea
in the dynamical compactification model of KK cosmology\cite{64},
i.e.,  the extra dimensions contract while the four visible
dimensions expand in order to assume that the present universe
satisfies\cite{57}
\begin{equation}\label{30}
a^{3}(t)\epsilon^{\frac{n}{2}}(t)=constant,
\end{equation}
where $n$ is a positive number. Hence, we have $\frac{\dot{\epsilon}}{\epsilon}=-\frac{6}{n}H$.
In addition, according to the spirits of the literature\cite{60},
we expect $\epsilon^{-\frac{1}{2}}\phi^{\frac{1}{m-1}}\sim1$, at least for the present period,
which results in $\phi_{0}=1$ with no dimension and
$\left(\frac{\dot{\phi}}{\phi}\right)_{t_{0}}=\frac{m-1}{2}\left(\frac{\dot{\epsilon}}{\epsilon}
\right)_{t_{0}}$. It follows that the initial conditions mentioned
above in summary are:
\begin{equation}\label{31}
a_{0}=\epsilon_{0}=\phi_{0}=1,\\
\dot{a}_{0}=H_{0},~~\dot{\epsilon}_{0}=-\frac{6}{n}H_{0},~~\dot{\phi}_{0}=\frac{3(1-m)}{n}H_{0}.
\end{equation}
By complicated calculations the 5D curvature scalar can be given as
\begin{equation}\label{32}
R=6\left(\frac{\ddot{a}}{a}+\frac{\dot{a}^{2}}{a^{2}}\right)+\frac{\ddot{\epsilon}}{\epsilon}
-\frac{1}{2}\left(\frac{\dot{\epsilon}}{\epsilon}\right)^{2}
+3\frac{\dot{a}}{a}\frac{\dot{\epsilon}}{\epsilon}.
\end{equation}
Substituting Eq.(\ref{27}) into (\ref{32}), we can obtain
\begin{equation}\label{33}
\left[\frac{1}{\alpha m\left(1-2\kappa\rho\right)}\right]^\frac{1}{m-1}=\frac{24m}{2m+1}\left
[1-q_{0}+\frac{3\left(1-m\right)}{n^{2}}\right]H^{2}_{0},
\end{equation}
where the deceleration parameter
$q\equiv-\frac{\ddot{a}a}{\dot{a}^{2}}=-\frac{1}{H^{2}}\frac{\ddot{a}}{a}$,
and $q_{0}$ is its present value. By virtual of (\ref{33}) and
 (\ref{29}), the relationship between parameters $m$ and $n$ can be
 obtained as follows
\begin{equation}\label{34}
6m^{2}(n-3)+m\{27+n[3+n(2-4q_{0})]\}+n^{2}(4q_{0}-5)-9=0.
\end{equation}

Similar to the above discussions, we find that in the reduced FGNMC
model the relationship between parameters $m$ and $n$ is the same as
Eq.(\ref{34}), but different from the one in the FGNC model, which is
given by
\begin{equation}\label{35}
4mn^{2}q_{0}-4n^{2}q_{0}+5n^{2}-2mn^{2}-18m+9=0.
\end{equation}
Evidently, it is just the same as Eq.(26) in Ref.\cite{60}. For the
sake of comparison we have listed the related quantities to the
FGCMC, FGNMC, FGNC and EG models in Table 1, from which it is easy
to find that there is no need to constrain the value of the
parameter $m$ on the Einstein gravity in the reduced 4D spacetime.

\begin{table}[!htb]
\centering \hspace{0.1cm}\small{\begin{tabular}{|c|c|c|c|c|c|c|} \hline
5D models &$\mathcal{L}^{(5)}$& $f_{1}(R)$ & $f_{2}(R)$ & $\phi\equiv f'_{1}+2\kappa Bf'_{2}$ &
$\dot{\epsilon}_{0}$ & $\dot{\phi}_{0}$ \\
\hline FGCMC &$\frac{1}{2\kappa}f_{1}+L_{m}f_{2}$ & $\alpha R^{m}$ & $\alpha R^{m}$ & $\alpha m(1-2
\kappa\rho)R^{m-1}$ &$-\frac{6}{n}H_{0}$&$\frac{3(1-m)}{n}H_{0}$\\
\hline FGNMC & $\frac{1}{2\kappa}f_{1}+L_{m}(1+\lambda f_{2})$ & $\alpha R^{m}$ & $\alpha R^{m}$ & $
\alpha m(1-2\kappa\lambda\rho)R^{m-1}$ &$-\frac{6}{n}H_{0}$&$\frac{3(1-m)}
{n}H_{0}$\\
\hline FGNC &$\frac{1}{2\kappa}f_{1}+L_{m}$ & $\alpha R^{m}$ & 1 & $\alpha m R^{m-1}$ &$-
\frac{6}{n}H_{0}$&$\frac{3}{n}H_{0}$\\
\hline EG &$\frac{1}{2\kappa}R+L_{m}$ & $R$ & 1 & 1 &$-\frac{6}{n}H_{0}$& 0\\
\hline
\end{tabular}}
\caption{\small{The related quantities to the FGCMC, FGNMC, FGNC and EG models.}} \label{T1}
\end{table}

As we knew, one necessary and sufficient condition for the present
accelerated expansion of the universe is that the deceleration
parameter satisfies the condition $q_{0}<0$, and according to the
present observation the allowed range for $q_{0}$ is
$q_{0}=(-0.57\pm0.10)$\cite{63}. Now we use this criteria to
determine the allowed ranges for the parameters $m$ and $n$. From
the equation (\ref{34}) it is not difficult to find that it has two
sets of solutions for positive and negative values of $m$,
respectively, but Eq.(\ref{35}) only has one due to $n>0$, which are
illustrated in Figs.1 and 2, respectively. From the figures, it is
evident to see that there do exist suitable values of the parameters
$m$ and $n$, which are just the shaded parts surrounded by the two
curves in Figs.1 and 2. It follows that in the reduced 4D spacetime
the present cosmic acceleration can be explained by our model rather
than dark energy.

\begin{figure}[!htb] \vspace{0.4cm} \hspace{-0.6cm}
\centering
\includegraphics[width=190pt,height=140pt]{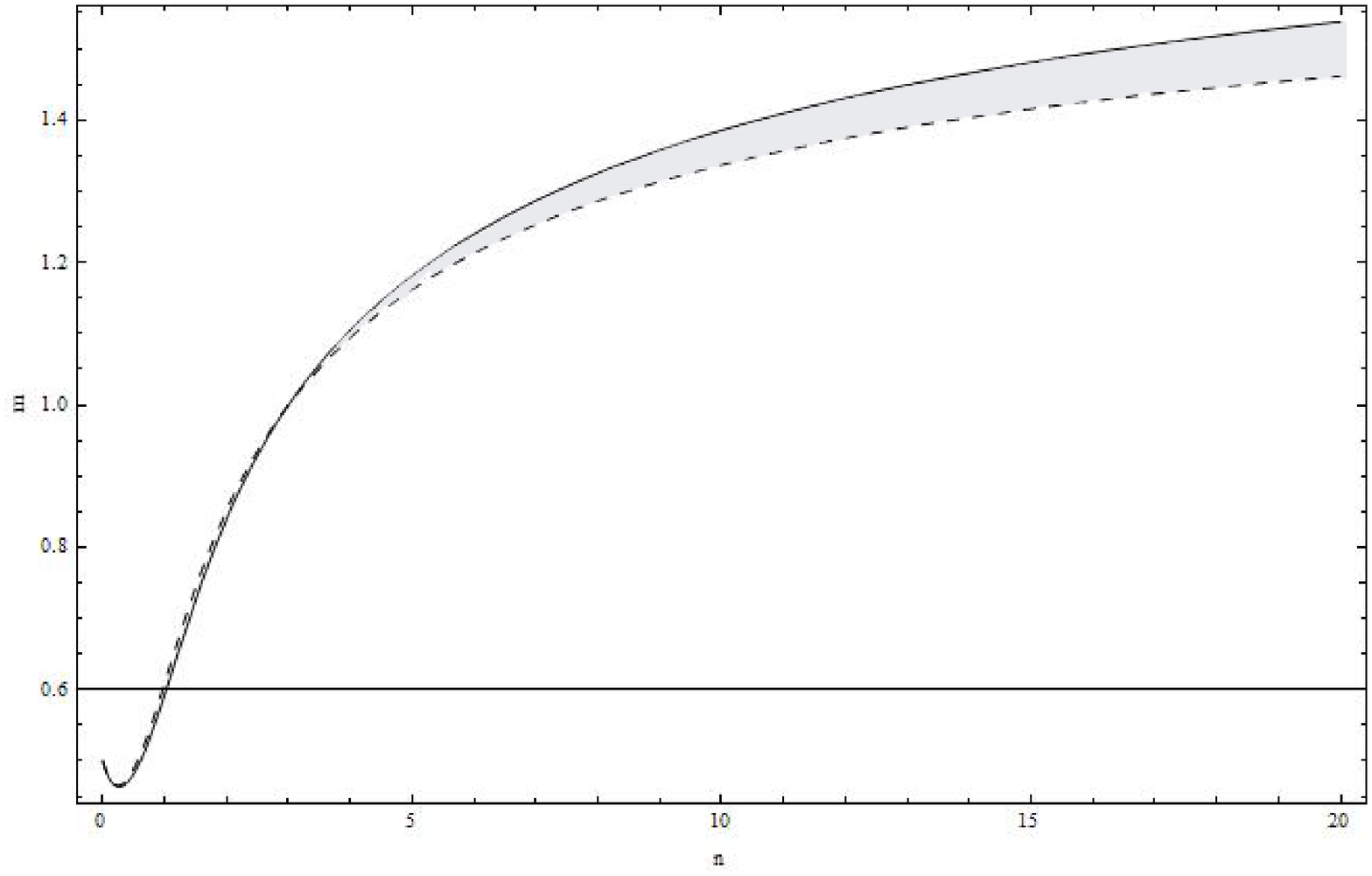} \hspace{-0cm}
\includegraphics[width=190pt,height=140pt]{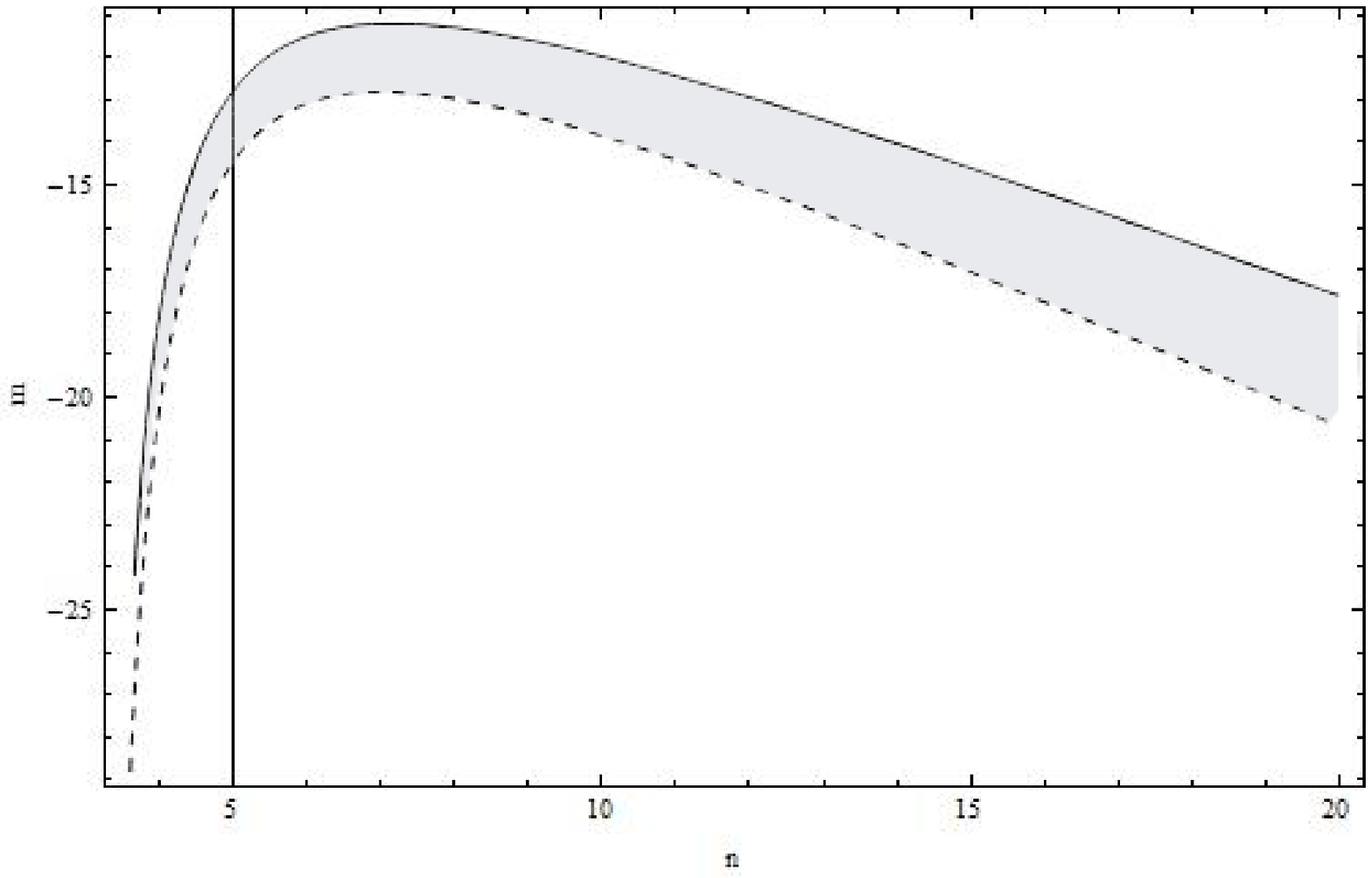} \hspace{-1.2cm}
\\~~~(a)~~~~~~~~~~~~~~~~~~~~~~~~~~~~~~~~~~~~~~~~~~~~(b)~~\\ \caption[Fig.1]
{\small{The constraints on the parameters $m$ and $n$ of FGCMC or
FGNMC models with $q_{0}\in(-0.67,-0.47)$. The solid line shows the
case of $q_{0}=-0.47$ and the dotted line shows the case of
$q_{0}=-0.67$ in each figure. (a) corresponds to the positive
solution of $m$ and (b) corresponds to the negative solution of
$m$.}} \label{F1}
\end{figure}

\begin{figure}[!htb] \vspace{0.4cm} \hspace{-0.6cm}
\centering
\includegraphics[width=190pt,height=140pt]{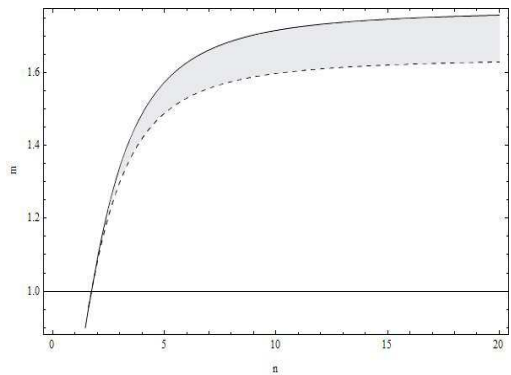} \hspace{-0cm}
\caption[Fig.2]
{\small{The constraints on the parameters $m$ and $n$ of FGNC model with $q_{0}\in(-0.67,-0.47)$.
The solid line shows the case of $q_{0}=-0.47$ and the dotted line shows
the case of $q_{0}=-0.67$.}} \label{F2}
\end{figure}

In addition, to illustrate the evolutional characters of some
cosmological quantities such as the scale factor $a(t)$, the
deceleration parameter $q(t)$ and the scalar field $\epsilon(t)$,
$\phi(t)$, around the present epoch in these models, we would here
choose some specific values of $m$ and $n$ in the admissible range,
for example, we choose $m=1.3$ with $n=8$, which results in
$\left[\frac{1}{\alpha
m\left(1-2\kappa\rho\right)}\right]^\frac{1}{m-1}\approx
13.75H_{0}^{2}$ in the reduced FGCMC model. Thus, by using
Eqs.(\ref{27})-(\ref{29}) and initial values, the evolutionary
trajectories of $a(t)$, $q(t)$, $\epsilon(t)$ and $\phi(t)$ can be
plotted in Figs.3 and 4.

\begin{figure}[!htb] \vspace{0.4cm} \hspace{-0.6cm}
\centering
\includegraphics[width=230pt,height=160pt]{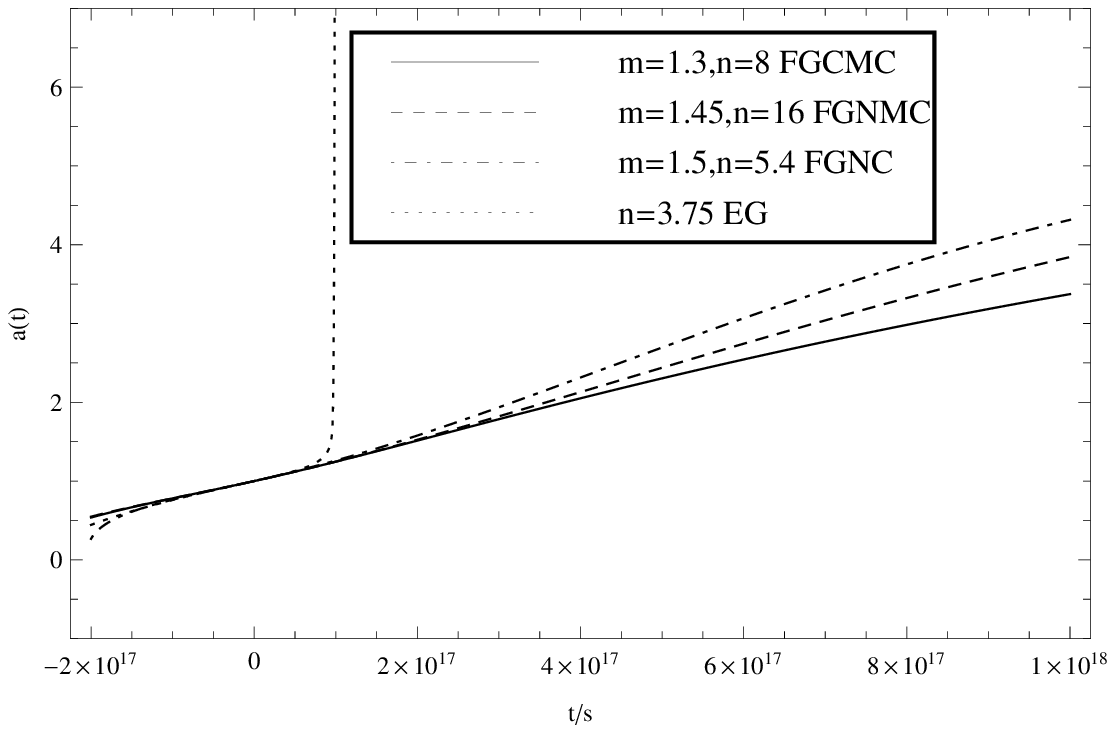} \hspace{-0cm}
\includegraphics[width=230pt,height=160pt]{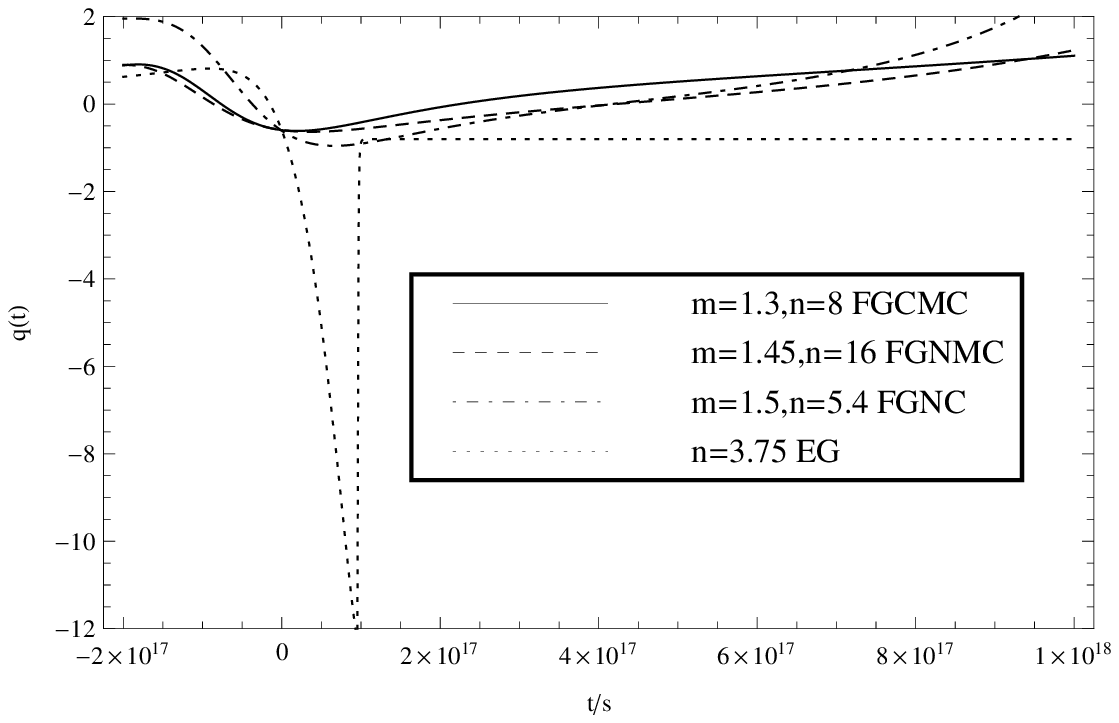} \hspace{-1.2cm}
\\~~~(a)~~~~~~~~~~~~~~~~~~~~~~~~~~~~~~~~~~~~~~~~~~~~(b)~~\\ \caption[Fig.3]
{\small{(a) and (b) respectively show the evolutionary trajectories
of the scale factor $a(t)$ and the deceleration parameter $q(t)$.
Here the current values $\rho_{0}=3.8\times10^{-28}kgm^{-3}$,
$H_{0}=2.3\times10^{-18}s^{-1}$ and $a_{0}=\epsilon_{0}=\phi_{0}=1$,
and $t=0$ corresponds to today.}} \label{F3}
\end{figure}

\begin{figure}[!htb] \vspace{0.4cm} \hspace{-0.6cm}
\centering
\includegraphics[width=230pt,height=160pt]{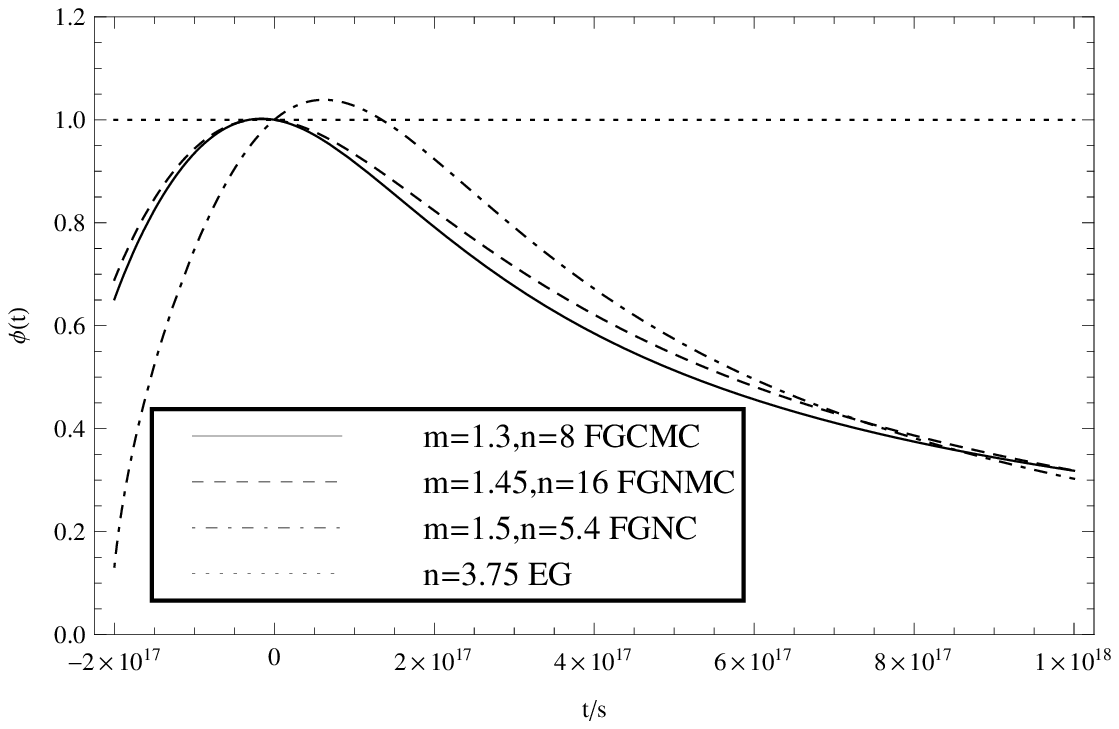} \hspace{-0cm}
\includegraphics[width=230pt,height=160pt]{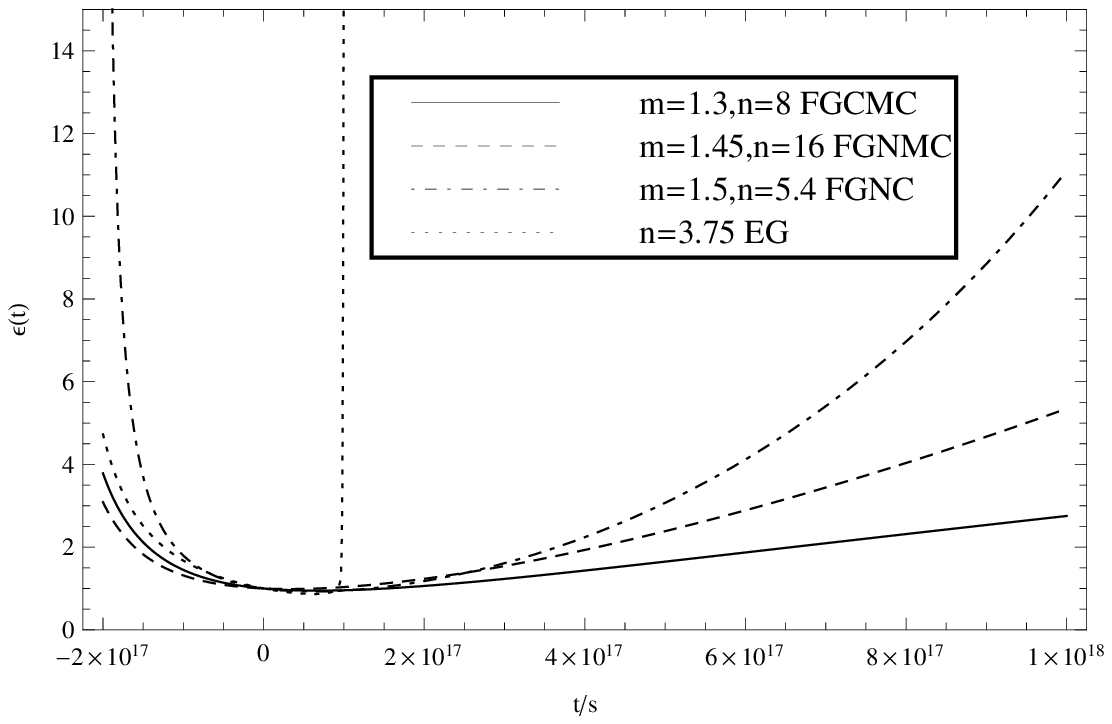} \hspace{-1.2cm}
\\~~~(a)~~~~~~~~~~~~~~~~~~~~~~~~~~~~~~~~~~~~~~~~~~~~(b)~~\\ \caption[Fig.4]
{\small{(a) and (b) respectively show the evolutionary trajectories
of the scalar field $\phi(t)$
 and $\epsilon(t)$ with the same parameters and current values
as those in Fig.3.}} \label{F4}
\end{figure}

Similarly, we also plot the evolutionary trajectories of the reduced
FGNMC, FGNC models as well as EG model in Figs.3 and 4. By analysis
we can find the present deceleration parameter $q_{0}=-0.6$, which
is consistent with observations\cite{63}. Moreover, the values of
the scalar factor $a(t)$ would be increasing slowly from the past to
the future in Fig.3(a), besides the one in the EG model has a rapid
increase at $t=(13.8+3.0)Gyr$ if the age of the universe is taken as
$13.8Gyr$\cite{5}; In Fig.3(b), each of the deceleration parameter
$q(t)$ in the FGCMC, FGNMC and FGNC models rolls from a positive
value to a negative one smoothly in the recent past. After each of the $q(t)$ reaches
the minimal value, it rolls back again and becomes positive, which
means that the universe will become decelerating in the future
rather than be endlessly accelerating. $q(t)$ in the EG model also
changes from a positive value to a negative one, but reaches the
minimal value sharply then has linear increase to a constant
acceleration, which suggests that the universe will be moving at a
constant acceleration; The scalar field $\phi(t)$ increases to a
maximum value in the past near the present in the FGCMC and FGNMC
models but to a maximum value for the present period in the FGNC
model. In addition, $\phi(t)$ is constant in the EG model;
$\epsilon(t)$ of the four models would decrease from a large value
in the past and increase slowly in the future in Fig.4, but the
severe change will happen once again at $t=(13.8+3.0)Gyr$ in the EG
model.

In general, the evolutionary trajectories of the scale factor $a(t)$, the deceleration parameter
$q(t)$ and the scalar field $\epsilon(t)$, $\phi(t)$ in the reduced
FGCMC, FGNMC, FGNC and EG models are able to explain the accelerated expansion of the universe
without introducing dark energy. Obviously, the results discussed in FGNC model are agree with Ref.\cite{60}.
However, the reduced EG model will be facing a dramatic change of the universe
in the next three billion years, and then be endlessly accelerating at a constant acceleration.

On the other hand, by taking the negative solutions of $m$ in Eq.(\ref{34}), we find that the
present value in the evolutionary trajectory of $q(t)$ is always not consistent with observations so that we
have to fail the negative solutions of $m$.

\section{Conclusions }

~~~~In this paper we have established the 5D generalized $f(R)$
gravity with curvature-matter coupling (FGCMC) by assuming that
there is a hypersurface-orthogonal spacelike Killing vector field in
the underlying 5D spacetime. This theory is the extension of the
original 4D generalized $f(R)$ gravity to the 5D spacetime, and it
is quite general and can give the corresponding results to the
Einstein gravity(EG), $f(R)$ gravity with both no-coupling(FGNC) and
non-minimal coupling(FGNMC) in 5D spacetime as special cases. It
follows that we have given the some new results besides previous
ones given by Ref.\cite{60}. Moreover, by using the approach of
Killing reduction the 5D generalized FGCMC can be reduced to the 4D
theory. Furthermore, in order to get some insight into the effects
of this theory on the 4D spacetime, by taking the specific type of
models with $f_{1}(R)=f_{2}(R)=\alpha R^{m}$ and
$B(L_{m})=L_{m}=-\rho$, not only have we listed the related
quantities corresponding to the different gravity models such as
FGCMC, FGNMC, FGNC as well as EG in Table 1, but also we have
illustrated the constraints on the model parameters $m$, $n$ and the
evolutionary trajectories of the scale factor $a(t)$, the
deceleration parameter $q(t)$ and the scalar field $\epsilon(t)$,
$\phi(t)$ in the reduced 4D spacetime in Figs. 1-2 and Figs. 3-4,
respectively. The research results show that there do exist suitable
values of the parameters $m$ and $n$, which means that the 5D
generalized FGCMC, FGNMC and FGNC can account for the present
accelerated expansion of the universe. Meanwhile, the evolutionary
laws of the scale factor $a(t)$, the deceleration parameter $q(t)$
and the scalar field $\epsilon(t)$, $\phi(t)$ in the reduced FGCMC,
FGNMC, FGNC and EG models are consistent with what are recognized.
These show the consistency between our results and the observations,
and would explain the current accelerated expansion of our universe
without introducing dark energy. However, the reduced EG model will
be facing a dramatic change of the universe in the next three
billion years, which would be endlessly accelerating at a constant
acceleration. Also, it is worth notice that in these models both
expansion and contraction of the extra dimension could result in the
present accelerated expansion of other spatial dimensions. Hence it
is reasonable to infer that the present accelerated expansion of
spatial dimensions could be certain basic character of the 5D
spacetimes.

\section*{Acknowledgements}
~~We are grateful to Dr. Song Li and Biao Huang for kind help and discussions.
The research work is supported by the National Natural
Science Foundation of China (Grant Nos.11175077, 11205078), the PhD Programs of
Ministry of Education of China (Grant No. 20122136110002)
and the Natural Science Foundation of Liaoning
Province, China (Grant Nos.20102124 and L2011189).

\end{document}